\documentclass[aps,preprint,amsmath,amssymb,t1,11pt]{revtex4}
\usepackage{graphicx}
\usepackage{epstopdf}
\usepackage{slashed}
\begin{document}
\title{Top quark anomalous FCNC production via $tqg$ couplings at FCC-hh}
\author{K.Y. Oyulmaz}
\email[]{kaan.yuksel.oyulmaz@cern.ch}
\author{A. Senol}
\email[]{senol_a@ibu.edu.tr} 
\author{H. Denizli}
\email[]{denizli_h@ibu.edu.tr}
\affiliation{Department of Physics, Bolu Abant Izzet Baysal University, 14280, Bolu, Turkey}
%\author{To be completed}
%\affiliation{...}
%\author{A. Yilmaz}
%\email[]{aliyilmaz@giresun.edu.tr}
%\affiliation{Department of Electrical and Electronics Engineering, Giresun University, 28200 Giresun, Turkey}
%\author{I. Turk Cakir}
%\email[]{ilkay.turk.cakir@cern.ch}
%\affiliation{Department of Energy Systems Engineering, Giresun University, 28200 Giresun, Turkey}
\author{O. Cakir}
\email[]{ocakir@science.ankara.edu.tr}
\affiliation{Department of Physics, Ankara University,  06100, Ankara, Turkey} 

\begin{abstract}
We investigate the top quark anomalous Flavor Changing Neutral Current (FCNC) $tqg$ interactions to probe limits on the couplings $\zeta_c$ and $\zeta_u$ through the $qg \to l\nu b$ signal suprocess at FCC-hh collider with center of mass energy of 100 TeV. To separate signal from relevant Standard Model background processes, selection criteria based on Boosted Decision Trees (BDT) is used with a set of useful kinematic variables. The sensitivities on the anomalous top FCNC couplings $\zeta_u$  and $\zeta_c$ are found to be $1.239\times10^{-4}$ and  $1.149\times10^{-4}$ for FCC-hh with $L_{int}$=10 ab$^{-1}$ at 95 \% C.L. including realistic detector effects of the FCC-hh baseline detector, respectively. The branchings $BR (t \to ug)$ and $BR (t \to c g)$ converted from obtained limits for FCNC couplings are at the order of $10^{-7}$ which is at least one order of magnitude better than the projected limits of HL-LHC with $L_{int}$= 3  ab$^{-1}$.
\end{abstract}
\maketitle

\section{Introduction}
The top quark being the most massive elementary particle in the Standard Model (SM) is an excellent probe not only to search the dynamics of electroweak symmetry breaking but also to test SM and Beyond the Standard Model (BSM) physics. The Flavour Changing Neutral Current (FCNC) interactions involving a top quark, other up type quarks ($u$, $c$) and neutral gauge bosons are forbidden at tree level and suppressed in loop level according to the Glashow-Illopoulos-Maiani (GIM) mechanism in the SM \cite{Glashow70}. Since predictions for SM branching ratios of the top quark FCNC decay to gluon, photon, $Z$ or Higgs boson and up-type quarks are out of range for current experimental sensitivities, the top quark FCNC interactions can have an important role to test new physics. In the BSM scenarios such as two-Higgs doublet model \cite{Eilam:1990zc}, supersymmetry \cite{Yang:1997dk}, technicolor \cite{Lu:2003yr}, and the minimal supersymmetric standard model \cite{Li:1993mg}, the branching ratios of top quark FCNC decays are predicted promisingly at the order of $10^{-6}$ - $10^{-5}$ due to enhancement on the production rate. 

Recent experimental results at 95\% confidence level (C.L.) on the top quark FCNC branching ratios for $t \to q g$, where $q$ indicates an up quark ($u$) or a charm quark ($c$),  are $BR$($t \to ug$) $< 4.0\times10^{-5}$ and $BR$($t \to cg$) $< 2.0\times10^{-4}$ reported by ATLAS Collaboration using $\sqrt s=$8 TeV data \cite{Aad:2015gea}, while $BR$($t \to ug$) $< 2.0\times10^{-5}$ and $BR$($t \to cg$)$ < 4.1\times10^{-4}$ are obtained by CMS Collaboration using combined $\sqrt s=$7 TeV and $\sqrt s=$8 TeV data \cite{Khachatryan:2016sib}. As a more promising and realistic project, high luminosity LHC (HL-LHC) expects to reach branching $BR$($t \to ug$)$ < 3.8\times10^{-6}$ ($9.8\times10^{-6}$) and $BR$($t \to cg$)  $< 32\times10^{-6}$ ($99\times10^{-6}$) for integrated luminosity of 3000 fb$^{-1}$ (300 fb$^{-1}$) at $\sqrt s=$14 TeV with a full simulation of the Phase-2 CMS detector upgrade \cite{CMS:2018kwi}. 

One can even expect to improve these limits at higher center of mass energies. The Future Circular Collider (FCC) project \cite{FCC} has great potential with an option of proton-proton (FCC-hh) collisions at 100 TeV center of mass energy with peak luminosity $5\times10^{34}$ $cm^{-2}s^{-1}$ \cite{Mangano:2017tke}. 

In this study, we investigate anomalous FCNC $tqg$ interactions to probe limits on couplings $\zeta_c$ and $\zeta_u$ couplings through the $qg \to l\nu b$ signal subrocess at FCC-hh collider. Realistic detector effects are included in the production of signal and background processes. Relevant SM backgrounds are considered and the sensitivity of anomalous FCNC $tqg$ couplings are searched by using multivariate analysis. Finally, the results are reported for different luminosity projections and compared with current experimental results.  
\section{FCNC $t\to qg$ vertices}
In the search of anomalous FCNC $tqg$ interactions at hadron colliders, the effective Lagrangian approach \cite{AguilarSaavedra:2008zc,AguilarSaavedra:2009mx} has been comprehensively studied in literature for hadron colliders \cite{Han:1998tp,delAguila:1999kfp,Belyaev:2001hf,Alan:2002wv,Cakir:2003cg,Yang:2004af,Cakir:2005rf,Zhang:2008yn,Cakir:2009rq,Drobnak:2010by,Gao:2011fx,Billur:2013ela,Agram:2013koa,Inan:2014mua,Koksal:2014hba,Hesari:2014eua,Khanpour:2014xla,Sun:2014qoa,Goldouzian:2014nha,Degrande:2014tta,Khatibi:2014via,Khatibi:2015aal,Hesari:2015oya,Sun:2016kek,Guo:2016kea,Liu:2016dag,Goldouzian:2016mrt,Zarnecki:2017cmf,Wang:2017pdg,Denizli:2017cfx,TurkCakir:2017rvu, Cakir:2018ruj,Shen:2018mlj,Liu:2018bxa,Banerjee:2018fsx,Chala:2018agk}. In this approach, FCNC interactions are described by higher-dimensional effective operators and added to four-dimensional SM Lagrangian. The FCNC Lagrangian of the $tqg$ interactions can be written as \cite{AguilarSaavedra:2008zc,AguilarSaavedra:2009mx}	
\begin{eqnarray}\label{eq1}
 L_{FCNC}&=& \frac{g_{s}}{m_{t}}\sum_{q=u,c} \bar q \lambda^a\sigma^{\mu\nu}(\zeta_{qt}^LP_L+\zeta_{qt}^RP_R)tG_{\mu\nu}^a + H.c.
\end{eqnarray}		
where $g_s$ is the strong coupling constant, $\lambda^a$ are the Gell-Mann matrices with $a=1,...,8$ and $\zeta_{qt}^{L(R)}$ is the strength of anomalous FCNC couplings for $tqg$ vertices; $P_{L(R)}$ denotes the left (right) handed projection operators. For the FCNC interactions, the tensor $\sigma^{\mu\nu}$  is defined as $\sigma^{\mu\nu}=\frac{i}{2}[\gamma^{\mu},\gamma^{\nu}]$. In this study, we assumed no specific chirality for the FCNC interaction vertices, i.e. $\zeta_{qt}^{L}=\zeta_{qt}^{R}=\zeta_q$ where $q$ denotes up or charm quark.		

\begin{figure}
\includegraphics[scale=1]{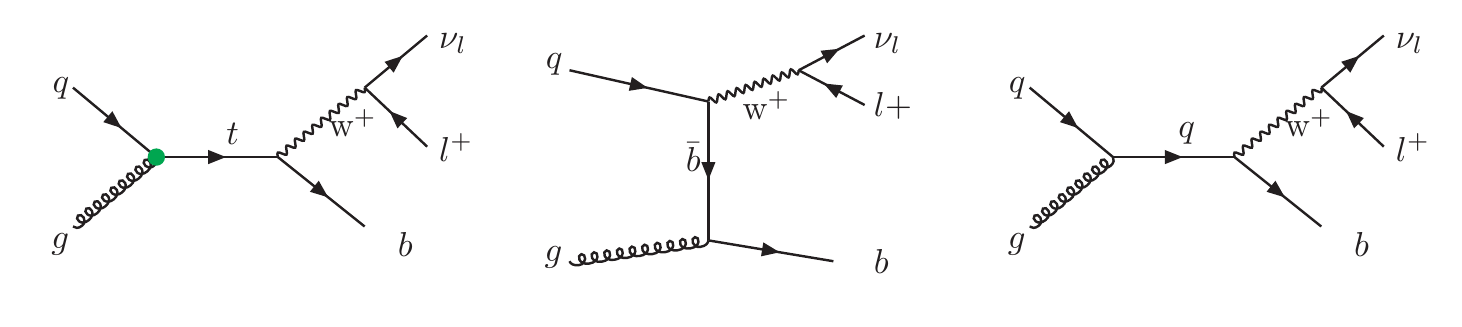}
\caption{  The Feynman diagrams for $q g\to l\nu b$ suprocess containing anomalous FCNC $tqg$ (green dot) and SM vertices.\label{fd}}
\end{figure}

Within the SM, top quarks are produced either as a pair via the strong interaction or the singly via weak interaction: i)  the $t$-channel process, ii) the $s$-channel process, and iii) the $Wt$ associated production at hadron colliders. With these production modes, the FCNC top-quark decays of $t \to qX$ mode with $X=H, Z,\gamma$, $g$ can be investigated through the final states of subsequent decays of particles. While the final states including $H$, $Z$, $\gamma$ can be searched promptly, the decay mode $t \to qg$ is almost indistinguishable from overwhelming backgrounds such as multijet-production via quantum chromodynamic (QCD) processes. To obtain better sensitivities for FCNC $tqg$ interactions, one can search direct top production, $qg \to t$, which originates from an up ($u$) or a charm ($c$) quark and gluon from in the initial state colliding hadrons and through subprocess combining immediately to form an s-channel top quark which then mostly decays to $Wb$. The Feynman diagrams of subprocess $qg \to l\nu b$  including the anomalous FCNC $tqg$ interactions and relevant SM background at tree level are shown in Fig.\ref{fd}.

\section{Signal and Background Simulation}
\begin{figure}
\includegraphics[scale=0.8]{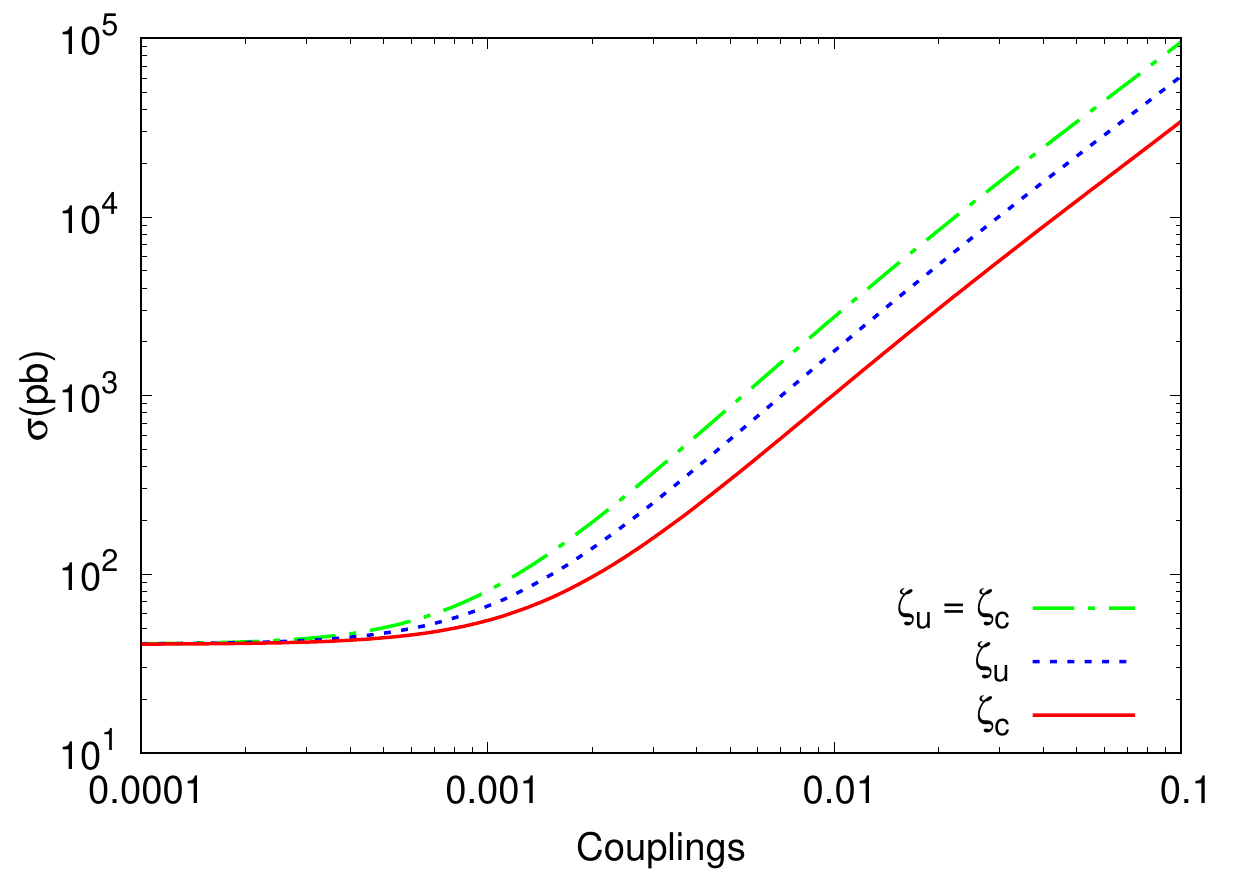}
\caption{  The cross section for the process $pp \to l\nu b+X$ including anomalous FCNC $tqg$ interactions with respect to $\zeta_c$ and $\zeta_u$ couplings.\label{cd}}
\end{figure}

 The effective Lagrangian in Eq.\ref{eq1} is defined in the FeynRules package as a Universal FeynRules Output (UFO) module \cite{Alloul:2013bka} and embedded into \verb|MadGraph2.5.3_aMC@NLO| \cite{Alwall:2014hca}. The cross sections at parton level for $qg \to l\nu b$ suprocess at 100 TeV center of mass energy have been evaluated with transverse momentum of the lepton $p_T^l>10$ GeV as a function of $\zeta_c$ and $\zeta_u$ couplings, which include signal  and interference between FCNC and SM as shown in Fig.\ref{cd}. As seen in Fig.\ref{cd}, noticeable deviations for the anomalous contributions starts around a coupling value $3\times10^{-4}$. Moreover, the contribution of $\zeta_u$ coupling is larger than the $\zeta_c$ coupling because of the dominant up quark parton distribution function at 100 TeV center of mass energy.

Since we study the FCNC $tqg$ couplings via $pp \to l\nu b$ process at FCC-hh, the final state topology of signal process consists of a charged lepton, missing energy and a b-tagged jet. The following relevant SM background processes having the same or similar final state topology are considered as backgrounds;
\[\begin{array}{clll}
\bullet&\mathrm{SM}:& pp\to l\nu b & \sigma= 4.060\times10^1~\mathrm{pb} \\
\bullet&Wj: &pp\to Wj &\sigma= 4.361\times10^5~\mathrm{pb} \\  
\bullet&Zj:& pp\to Zj' &\sigma= 1.583\times 10^5~\mathrm{pb} \\
\bullet&WW&pp\to W W &\sigma= 6.588\times 10^2~\mathrm{pb} \\  
\bullet&WZ&pp\to W Z &\sigma= 2.500\times 10^2 ~\mathrm{pb} \\ 
\bullet&ZZ&pp\to Z Z &\sigma= 9.990\times 10^1 ~\mathrm{pb} \\ 
\bullet&tt:& pp\to t\bar t &\sigma= 2.533\times 10^4~\mathrm{pb} \\ 
\bullet&tW:& pp\to t W & \sigma= 1.190\times 10^3 ~\mathrm{pb} \\  
\bullet&Wbb& pp\to  W b b & \sigma= 3.874\times 10^{2} ~\mathrm{pb} \\
\bullet&Wbj& pp\to  W b j & \sigma= 5.845\times 10^{3} ~\mathrm{pb} \\
\bullet&Wjj:& pp\to  W j j & \sigma= 2.635\times 10^5 ~\mathrm{pb} \\
\bullet&Zjj:& pp\to  Z j j & \sigma= 8.599\times 10^4 ~\mathrm{pb} \\
\end{array}
\]	 
where $j=u,\bar u, d, \bar d, s, \bar s, c, \bar c, g $ and $j' = j, b, \bar b$. The cross sections for each background processes have been computed at the resonance level with $p_T^l>10$ GeV and $p_T^j>20$ GeV cuts. Here, the SM refers to SM background of the same final state with the signal process shown the last two SM Feynman diagrams in Fig.\ref{fd}. The $Wj$ is considered as background candidate due to any misidentification of light quark as a mistagged b-jet in detector when $W$ boson decays leptonically. The $Zj'$ process is considered to another relevant background in this study. The diboson backgrounds; $WW$, $WZ$ and $ZZ$ are also included as a background with final states having either one or two charged leptons.
%\textbf{In a hadron collider, the top quarks can be produced at leading order via two ways in the SM framework: in pairs via strong interaction or in association with a jet or a W boson via weak interactions, also known as single. The production of single-top quark occurs via three different modes: the s-channel involving time-like W, the t-channel involving space-like W, and the tW production involving an on-shell W boson.} 
Since $Wj$ decay modes of top quark are suppressed relative to $Wb$ by the square of the CKM matrix-elements in the SM, the top quark decays almost exclusively to a bottom quark and a $W$ boson. The W decays to quark-antiquark pair (hadronic) or to a charged lepton and a neutrino (leptonic). So, the pair production of top quark and $tW$ events can be classified through the decays of the $W$ bosons as semileptonic, doubly hadronic and doubly leptonic. We consider the pair production of top quark ($tt$) having two b-tagged jets, opposite-charged leptons, and a (large) missing transverse energy and the $tW$ process characterized by the presence of a $b$-tagged jet, two isolated leptons with opposite charge, and a substantial amount of missing transverse energy due to the presence of the neutrinos in the doubly leptonic final states as backgrounds. The $Wbj$ (one $b$-tagged jet, one isolated leptons, and a missing transverse energy and a jet) and $Wbb$ (two $b$-tagged jet, one isolated leptons, and a missing transverse energy) production processes including off-shell top quarks are also taken into account as background for semileptonic final state. Finally, $Wjj$ and $Zjj$ background processes are included in the analysis considering leptonic decay channels of $W$ and $Z$ bosons. In the process of obtaining background estimations, jet matching procedure has not been considered.

The 10$^{6}$ events are generated by using \verb|MadGraph2.6.3.2aMC@NLO| for each signal and background processes with NNPDF23L01 \cite{Ball:2012cx} parton distribution functions and the renormalization and factorization scales are set to the $M_Z=91.188$ GeV. The  PYTHIA8.2  \cite{Sjostrand:2014zea} is utilized in parton showering and hadronization of generated signal and background events. All resonances $t$, $W$ and $Z$ decayed in PYTHIA8.2. Produced jets inside the events are clustered using FastJet3.2.1\cite{Cacciari:2011ma} with anti-kt algorithm \cite{Cacciari:2008gp} where a cone radius $R = 0.4$ and $p_T^{min}=30$ GeV. FCC-hh detector card embedded into Delphes 3.4.1 \cite{deFavereau:2013fsa} is used to include realistic detector effects of the FCC-hh baseline detector.
\begin{table}
\caption{Event selection and kinematic cuts used for signal and background events for BDT training. \label{cut}}
\begin{ruledtabular}
\begin{tabular}{lccc}
Preselection & $N_l \geqslant 1$ and $N_b \geqslant 1$ \\
Kinematic &  $p_{T}^{l} > 20$ GeV, $p_{T}^b > 30$ GeV\\
& $|\eta^{b}|<2.5$,$|\eta^{l}|<2.5$   \\
& MET $> 20$ GeV , $\Delta R (l,b) >$ 0.7  \\
W  Transverse mass & 45 GeV $< m_{T}^W < $90 GeV \\
Top Transverse mass & 100 GeV $< m_T^{top} <$ 200 GeV  \\
\end{tabular}
\end{ruledtabular}
\end{table}

\section{Event Selection}
\begin{figure}
\includegraphics[width=15cm, height=18cm]{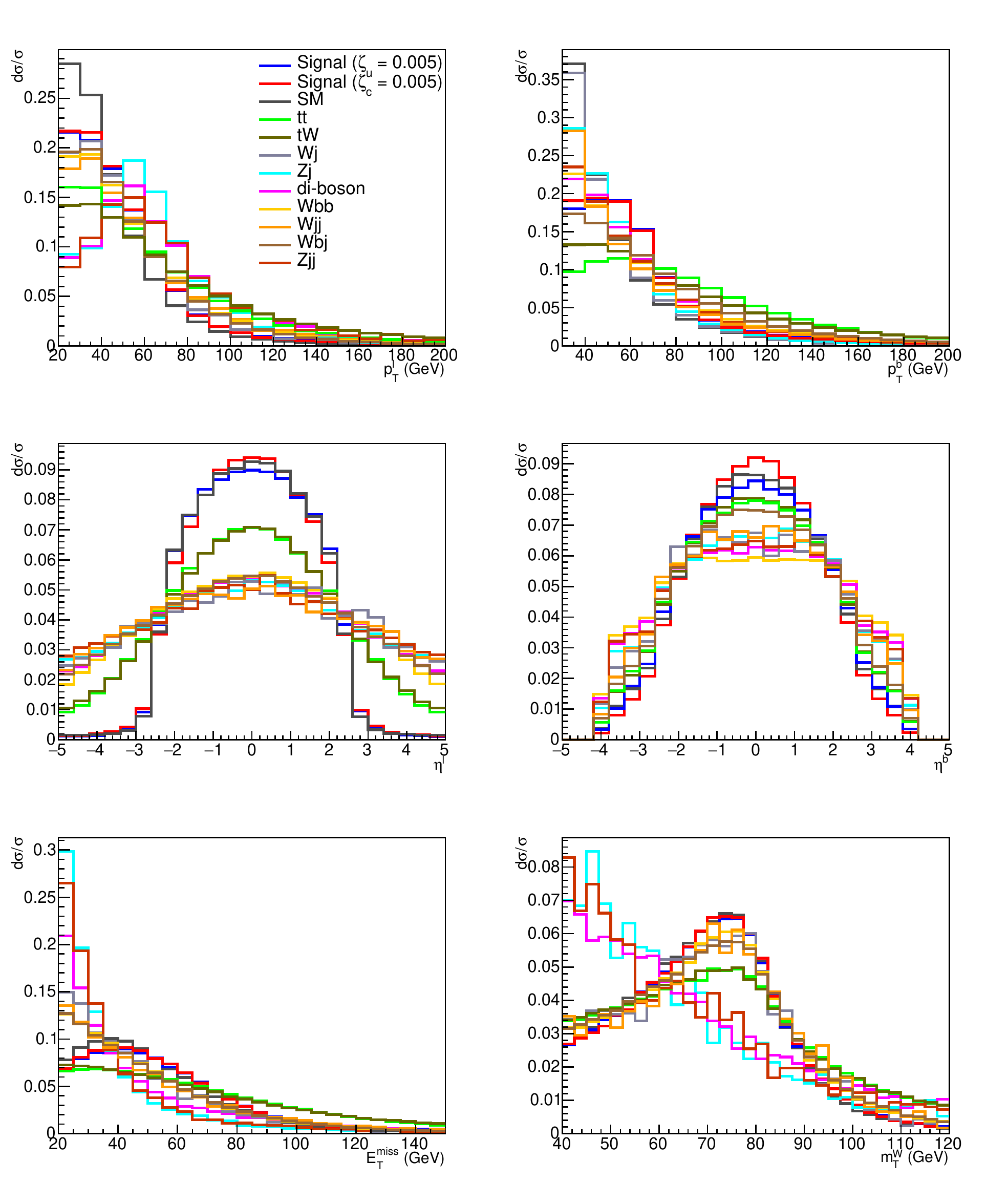}
\caption{Comparison of kinematic distributions of the final state particles for signal ($\zeta_u = 0.005$ and $\zeta_c = 0.005$ couplings) and all relevant backgrounds after requiring at least one lepton and b-tagged jet. These distributions are normalized to one. \label{kinematic}}
\end{figure}

\begin{figure}
\includegraphics[width=17cm, height=7cm]{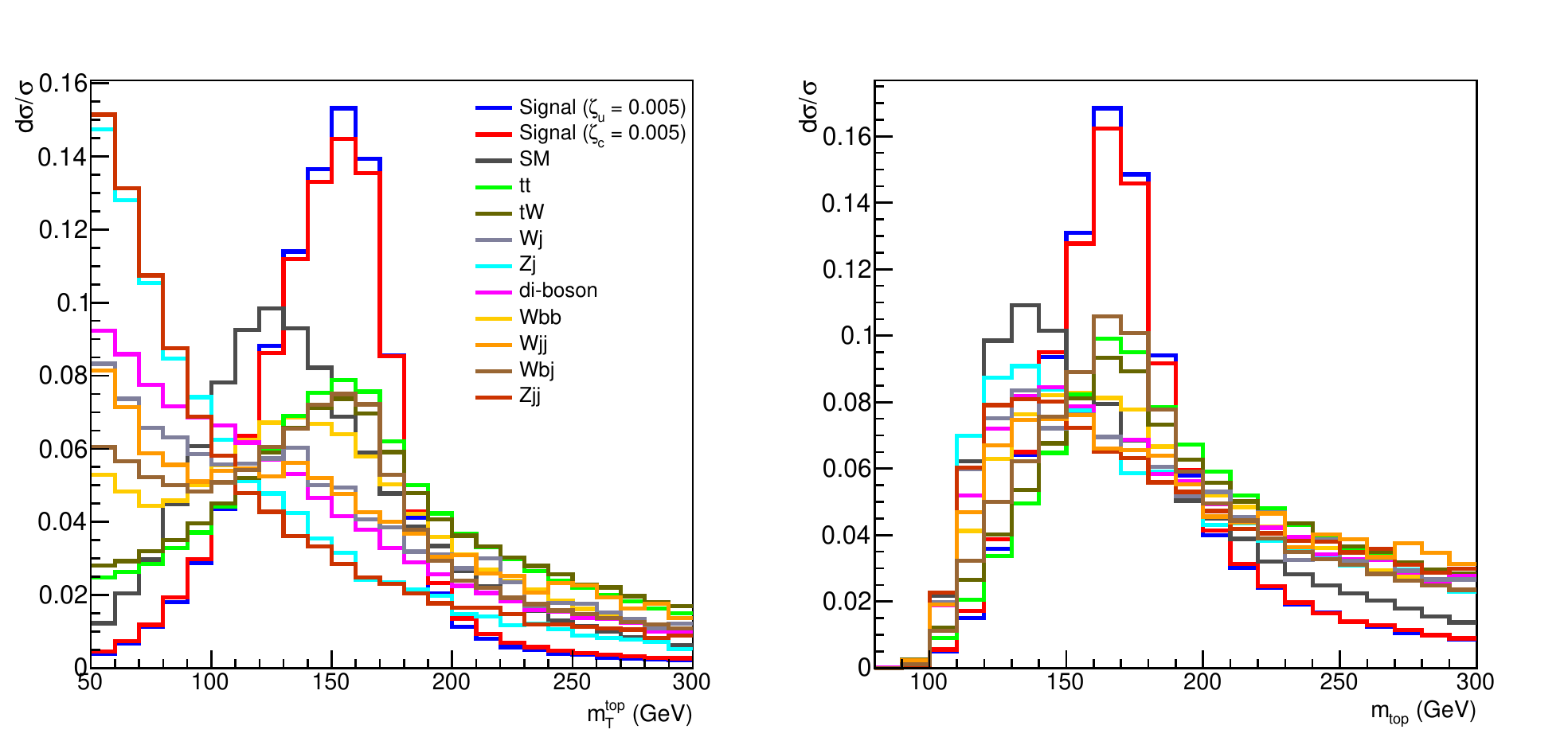}
\caption{ Comparison of the transverse mass ($m_T^{top}$) and invariant mass ($m_{top}$) distributions of reconstructed top quark for signal ($\zeta_c = 0.005$ and $\zeta_u = 0.005$ couplings) and all relevant backgrounds after requiring at least one lepton and b-tagged jet. These distributions are normalized to one.\label{tRec}}
\end{figure}

\begin{figure}
\includegraphics[width=17cm, height=5cm]{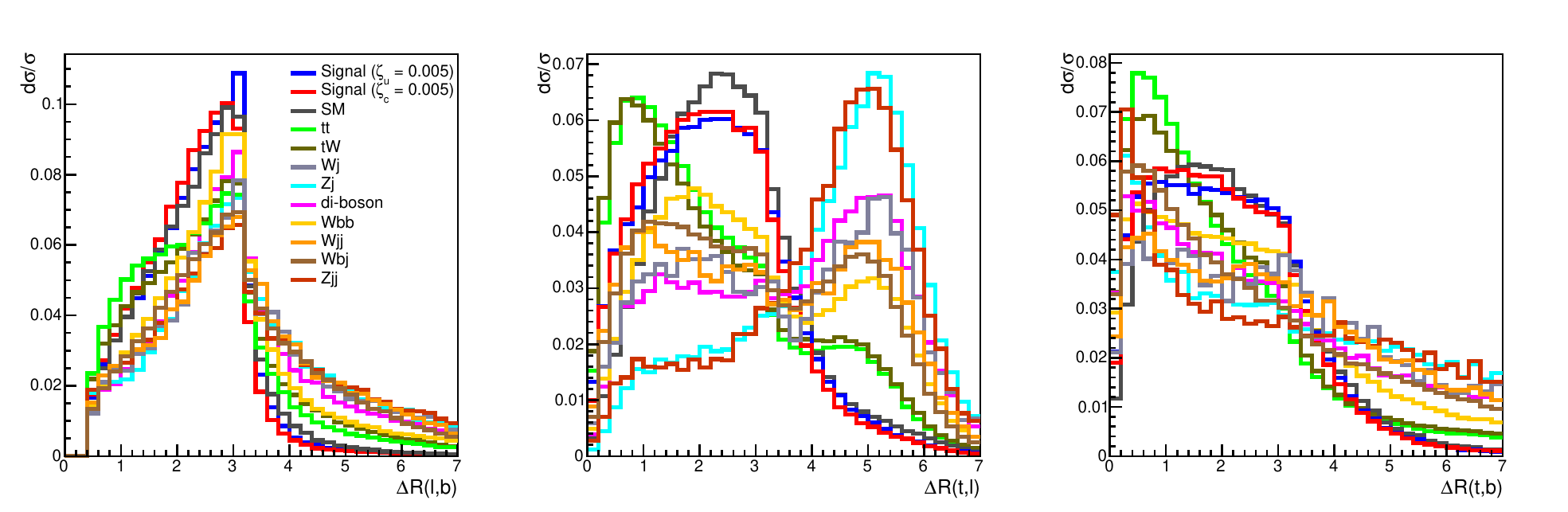}
\caption{ The distance between particles in pseudorapidity-azimuthal angle plane for leading lepton and b-jet $\Delta R (l,b)$, recontructed top and leading lepton $\Delta R (t,l)$ as well as leading b-jet $\Delta R (t,b)$, respectively. These distributions are normalized to one.\label{deltaR}}
\end{figure}

Characteristic signature of the $qg \to l\nu b$ signal process suggests to work with events having at least one isolated lepton (electron or muon), missing transverse energy and at least one jet which is required to be identified as a jet originating from $b$-quark for the analysis. Distributions of transverse momentum and pseudorapidity of leading lepton and b-tagged jet for signal and all relevant backgrounds are given in the first and second row of the Fig. \ref{kinematic}, respectively. Due to existence of neutrino in the final state of signal events, $W$-boson transverse mass is reconstructed as $m_T^W = \sqrt{2(p_{T,l} E_T^{miss} -  \overrightarrow p_{T,l} \cdot \overrightarrow{E}_T^{miss})}$. Both missing transverse energy and reconstructed $W$-boson transverse mass are also swon at the bottom set of Fig. \ref{kinematic}. Reconstructing $W$ boson that decays leptonically has difficulty due to unknown neutrino momentum vector. In a hadron collider, missing transverse energy is the only variable to be measured and considered to be transverse energy of neutrino in a good approximation. Taking the $W$ boson mass constraint, longitudinal component of the neutrino momentum is given by
\begin{eqnarray}
p_{z,\nu}^{\pm}&=&\frac{1}{p_{T,l}^2} \Big( \Lambda p_{z,l}\pm \sqrt{\Lambda^2p_{z,l}^2-p_{T,l}^2(E_l^2 (E_T^{miss})^2-\Lambda^2)} \Big)
\end{eqnarray}
\begin{eqnarray}
\Lambda=(m_W^2/2)+\vec{p}_{T,l}\cdot \slashed{\vec p}_T
\end{eqnarray}
where the $E_l$, $p_{T,l}$ and $p_{z,l}$ are the energy, transverse and longitudinal momentum components of the leading lepton, respectively. The solution that gives the smallest absolute value chosen and other solutions are discarded in our study as in Ref. \cite{Belyaev:1998dn}. 
%Constructing the 4-momentum of neutrino leads to reconstruction of invariant mass of $W$ ($m_W$) and then transverse mass ($m_T^{top}$) as well as invariant mass of top quark ($m_{top}$). 
The reconstructed $m_T^{top}$ and $m_{top}$ distributions are presented in Fig. \ref{tRec}. As shown in Fig. \ref{tRec}, signal peaks in the actual mass region of the top quark. The distance between leading lepton and b-tagged jet is $\Delta R (l,b) = \sqrt{(\Delta \phi_{l,b})^2 + (\Delta \eta_{l,b})^2}$ where $\phi_{l,b}$ and $\eta_{l,b}$ are azimuthal angle and the pseudorapidity difference between leading lepton and b-tagged jet. Similarly, one can obtain the distance between reconstructed top and leading lepton $\Delta R (t,l)$, reconstructed top and b-tagged jet $\Delta R (t,b)$.The distribution of $\Delta R (l,b)$, $\Delta R (t,l)$ and $\Delta R (t,b)$ are shown in Fig. \ref{deltaR}. 
\begin{table}
\caption{The list of selected kinematical and reconstructed variables to be used in BDT.   \label{BDT}}
\begin{ruledtabular}
\begin{tabular}{lll}
Variable & Definition \\ \hline 
%$N_{j}$&Number of jets in the event \\
$N_{b}$&Number of b-tagged jets in the event \\
$N_{l}$&Number of leptons in the event \\
$p_{T}^{l}$&Transverse momentum of the leading lepton \\
$p_{T}^{b}$&Transverse momentum of the leading b-jet \\
$\eta^{l}$&Pseudorapidity of the leading lepton \\
$\eta^{b}$&Pseudorapidity of the leading b-jet \\
$E_T^{miss}$&Missing transverse energy \\
$m_T^W$&Transverse mass of the reconstructed $W$-boson \\
$m_W$&Invariant mass of the reconstructed $W$-boson \\
$p_{T}^{W}$&Transverse momentum of the reconstructed $W$-boson \\
$m_T^{top}$&Transverse mass of reconstructed top quark \\
$m_{top}$&Invariant mass of reconstructed top quark \\
$cos\theta$&Opening angle of three-vectors between leading lepton and leading b-jet  \\
$\Delta R (l,b)$&Distance between lepton and b-jet in $\eta$-$\phi$ plane \\
$\Delta R (t,l)$&Distance between reconstructed top quark and lepton in $\eta$-$\phi$ plane \\
$\Delta R (t,b)$&Distance between reconstructed top quark and b-jet in $\eta$-$\phi$ plane \\

\end{tabular}
\end{ruledtabular}
\end{table}

\begin{table}
\caption{The cross sections of the signal and relevant backgrounds after BDT analysis.   \label{cs_afterBDT}}
\begin{ruledtabular}
\begin{tabular}{lll}
Process &cross section (pb) \\ \hline 
Signal ( $\zeta_u$=0.001)  & $1.228 \times 10^{0}$ \\
Signal ( $\zeta_c$=0.001) & $1.032 \times 10^{0}$ \\
Signal ( $\zeta_u$=0.005) & $1.559\times10^{1}$ \\
Signal ( $\zeta_c$=0.005) & $1.001\times10^{1}$ \\
SM& $6.643\times10^{-1}$ \\
$Wj$ &$5.017\times10^{1}$ \\  
$tt$ & $1.992\times 10^{2}$ \\ 
$Zj$ &$6.490\times 10^{0}$ \\
$tW$ & $9.694\times 10^{0}$  \\  
$WW$ & $2.912\times 10^{-1}$ \\  
$WZ$ &$2.043\times 10^{-1}$ \\ 
$ZZ$ & $1.459\times 10^{1}$  \\ 
$Wbb$ & $2.272\times 10^{0}$  \\ 
$Wbj$ & $3.653\times 10^{1}$  \\ 
$Wjj$ & $5.217\times 10^{1}$  \\ 
$Zjj$ & $6.879\times 10^{-1}$  \\ 
\end{tabular}
\end{ruledtabular}
\end{table}

\section{Analysis}
\begin{figure}
\includegraphics[width=8cm, height=5cm]{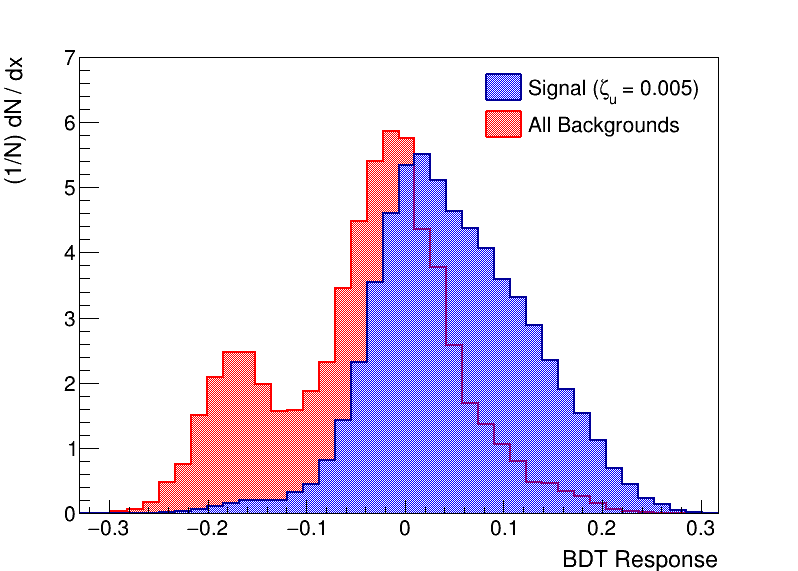}
\includegraphics[width=8cm, height=5cm]{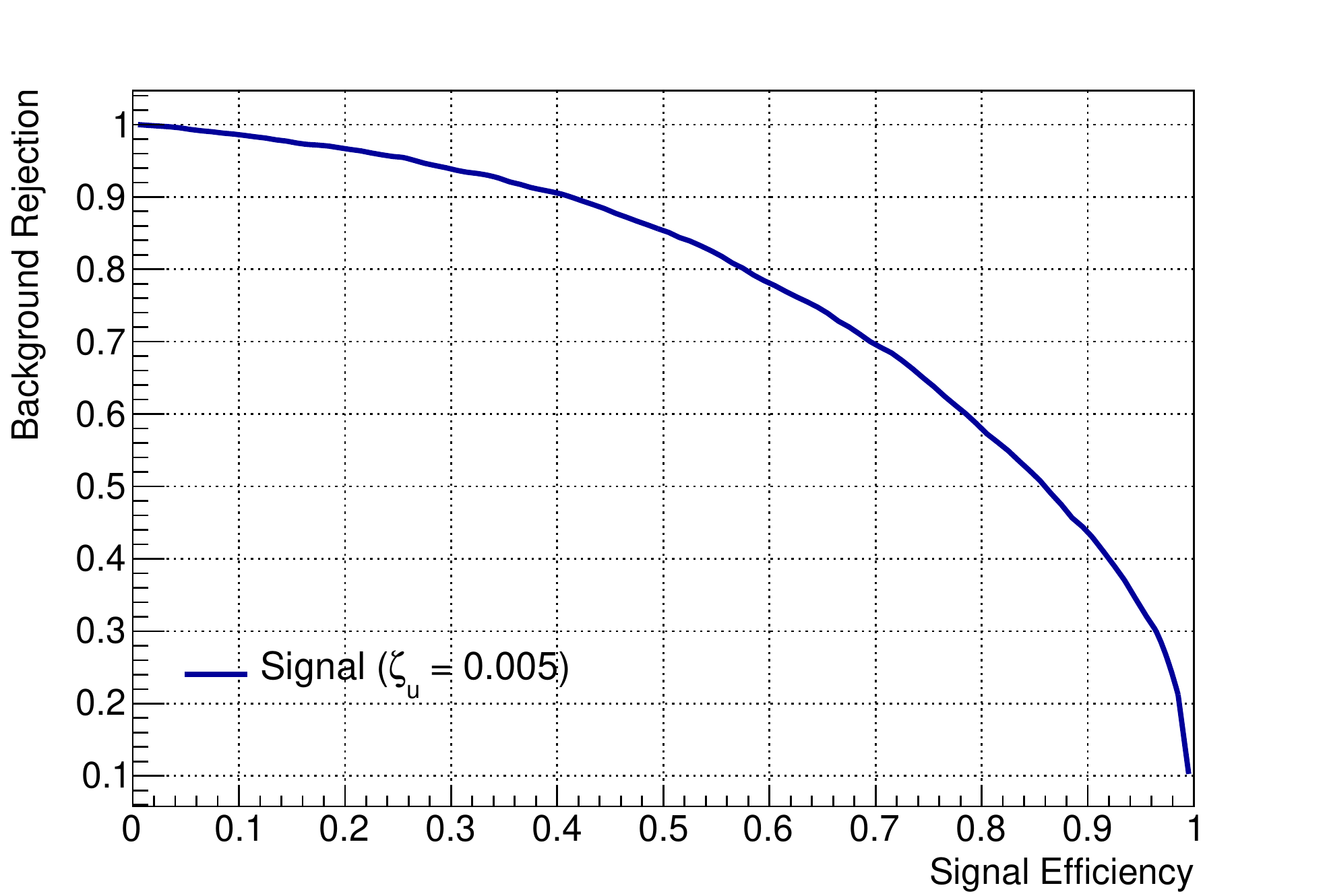}
\includegraphics[width=8cm, height=5cm]{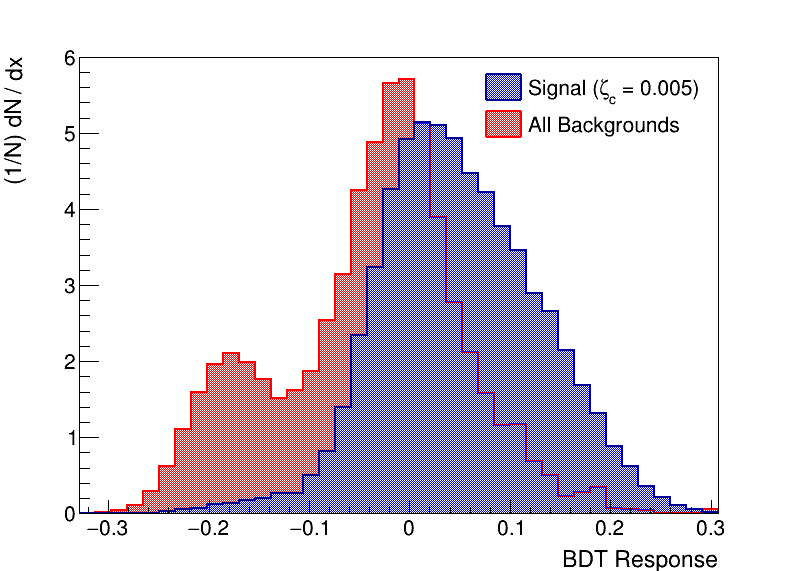}
\includegraphics[width=8cm, height=5cm]{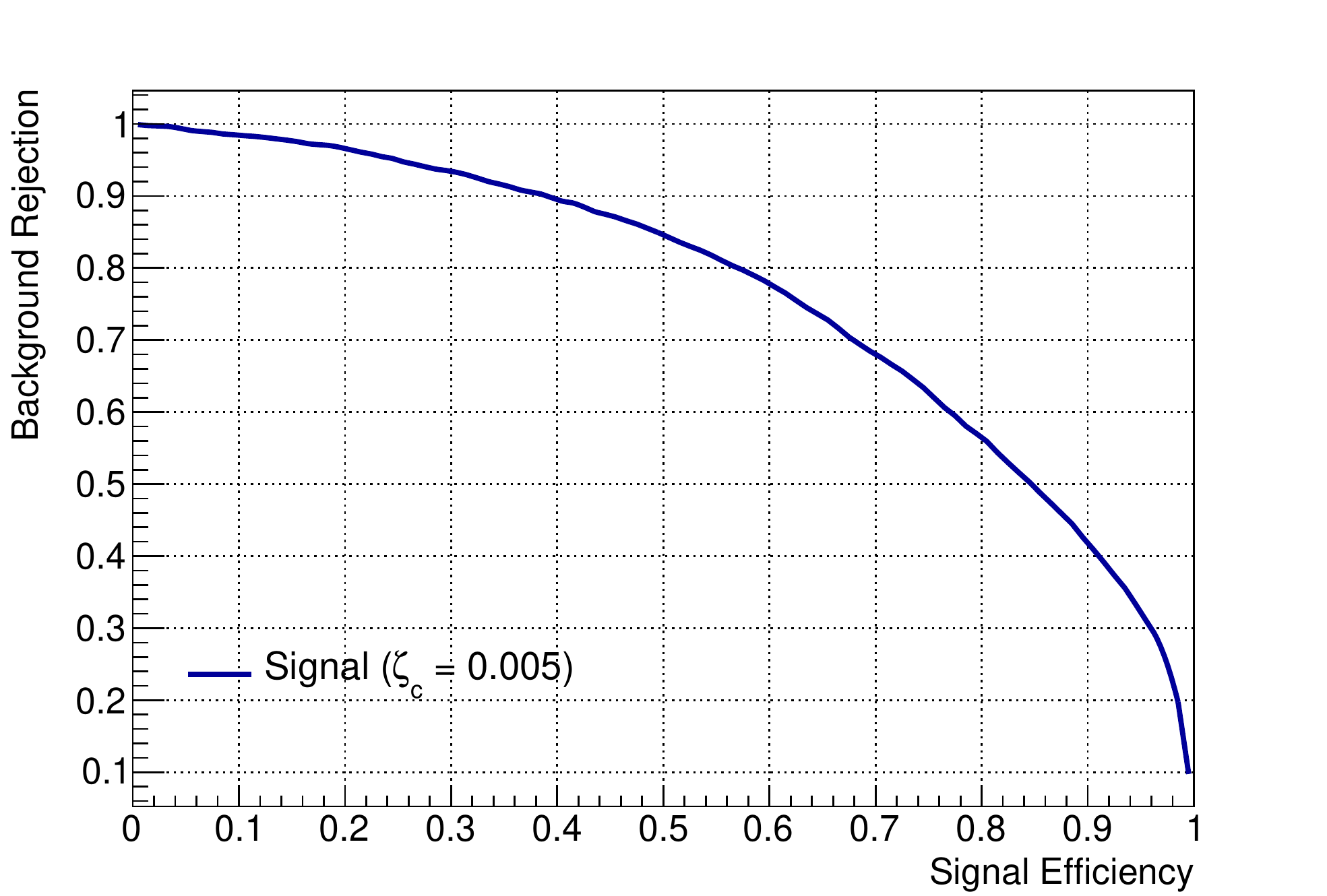}

\caption{ The distribution of the BDT response (on the left column) and Receiver Operating Characteristic (ROC) curve of the BDT (on the right column) for signal (couplings $\zeta_c = 0.005$ and $\zeta_u = 0.005$ couplings) and all relevant backgrounds. \label{BDT_output}}
\end{figure}

\begin{figure}
\includegraphics[width=8cm, height=7cm]{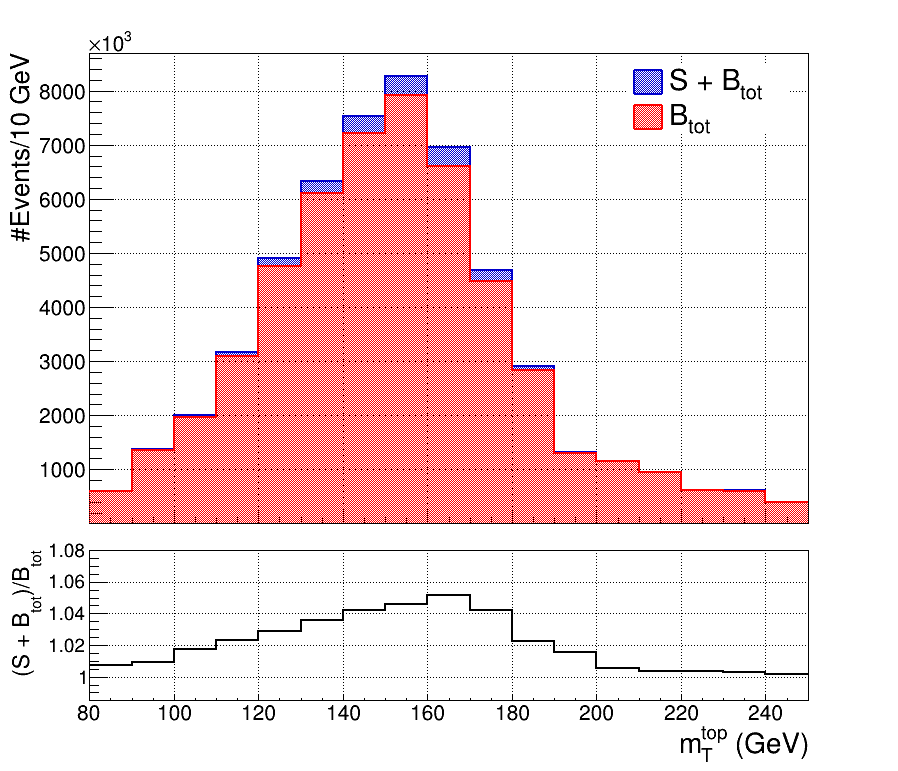}
\includegraphics[width=8cm, height=7cm]{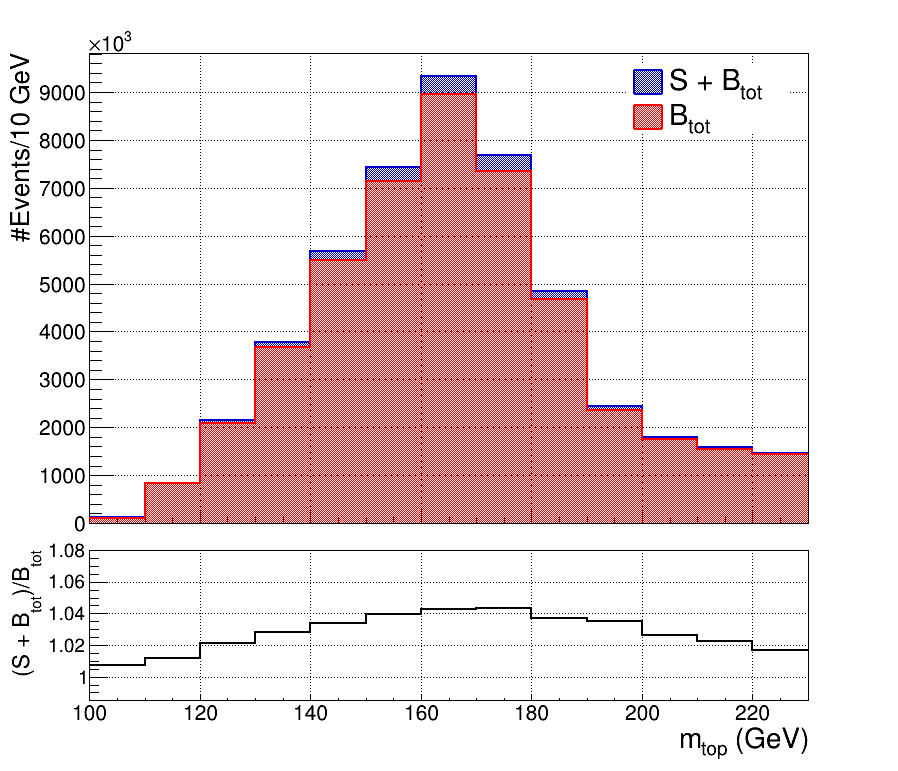}

\caption{The reconstructed transverse mass (on the left) and invariant mass (on the right) distributions of the top quark for signal ( couplings $\zeta_u = 0.005$ and $\zeta_c = 0.005$ couplings) and relevant SM background processes with an optimum BDT cut value 0.008. These distributions are normalized to $L_{int}$=100 fb$^{-1}$\label{mtop_inv}}
\end{figure}

Separation of signal from background events have been carried out by a multivariate technique \cite{Hocker:2007ht,Therhaag:2009dp} in the Toolkit for Multivariate Analysis (TMVA), particularly Boosted Decision Trees (BDT). Event selection cuts are defined from kinematic distributions given in Figs.  \ref{kinematic}-\ref{deltaR} and  summarized in Table \ref{cut}. Since the choice of input variables is important to train the BDT, we include total of 16 variables; lepton properties, b-tagged jet properties, missing transverse energy, invariant masses, and other variables which help to promote signal over the background. The topology of the signal process does not require light jet in the final state. Therefore we did not veto light jets in the BDT analysis, that is, the number of light jets and  kinematic variables such as $p_T$, $\eta$, $\Delta R(b,j)$ and  $\Delta R(l,j)$ are not considered in the BDT input variable list. A detailed list of the variables is given in Table \ref{BDT}. From these variables, the transverse mass of top quark is selected as a target in BDT. The 50\% of events is used in the test while the other half is used in the training for signal and background multivariate analysis. Distribution of the BDT classifier response for the considered signal  and total background events is shown in Fig.\ref{BDT_output}.  Top left plot in Fig.\ref{BDT_output} corresponds to BDT response for signal $\zeta_u$ =0.005 and all the other overwhelming backgrounds for trained samples. On the other hand, top right plot so-called Receiver Operating Characteristic (ROC) curve shows the signal efficiency as a function of background rejection. Bottom two plots in Fig.\ref{BDT_output} are BDT response and ROC curve for signal $\zeta_c$=0.005 and all other backgrounds. As it can be seen from this figure, performance of BDT is quite well and signal can be separated from backgrounds. Therefore one can determine optimal cut on reconstructed BDT distributions considering signal efficiency. In our study, 70\% signal efficiency has been taken into account in the determination of optimal BDT cut which varies for each signal scenarios with different values of couplings $\zeta_c$ and $\zeta_u$.  Applying the optimal BDT cut value to signal and background events, transverse mass (on the left) and invariant mass (on the right) distributions of reconstructed top quark are shown in Fig.\ref{mtop_inv}. The cross sections of the signal for two benchmark values of anomalous couplings ($\zeta_q$=0.001 and $\zeta_q=0.005$) and relevant backgrounds processes are computed with the optimal BDT cut value and tabulated in Table~\ref{cs_afterBDT}. The cross section of a given background process are large at the resonance level in full phase space, but small after cuts and b-tagging have been applied in the final state as seen in Table~\ref{cs_afterBDT}.

\begin{figure}
\includegraphics[width=8cm, height=7cm]{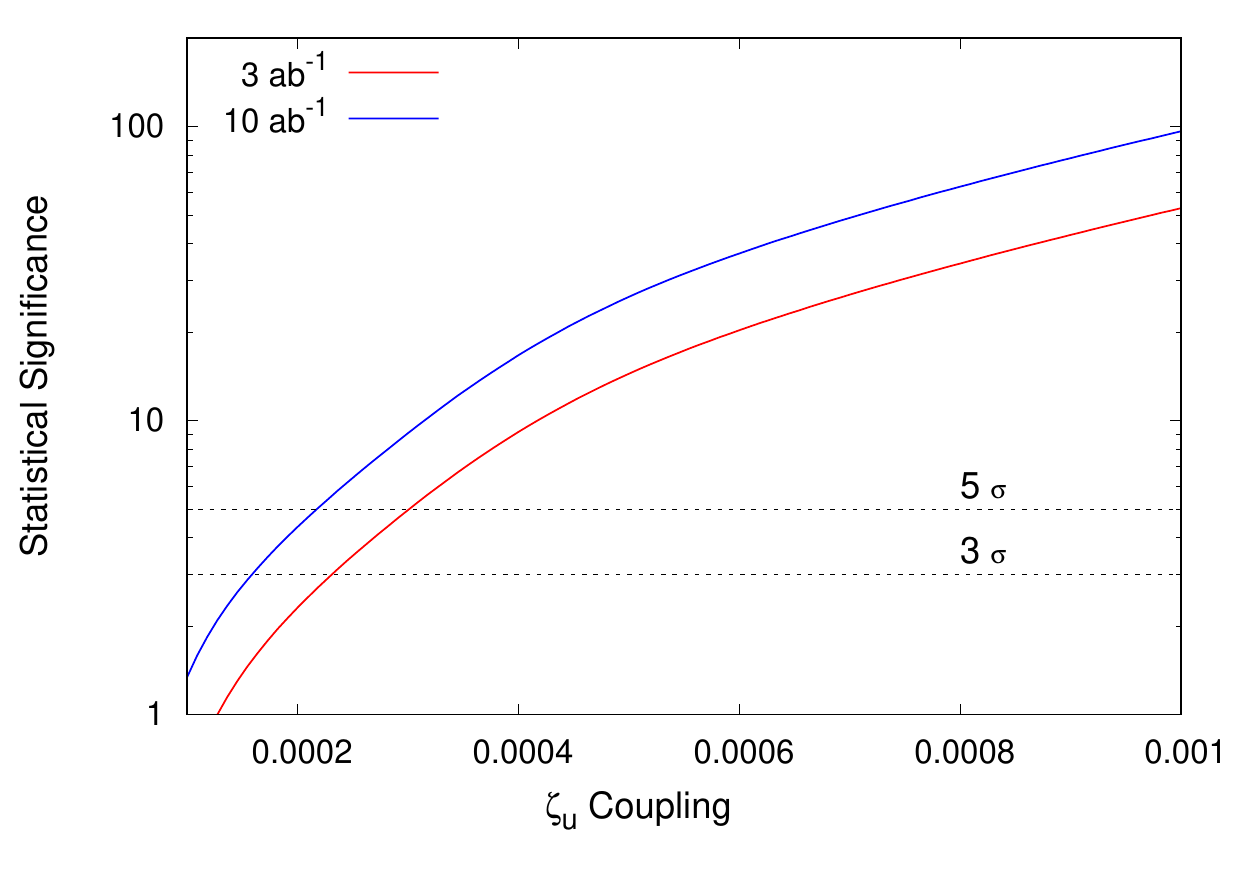}
\includegraphics[width=8cm, height=7cm]{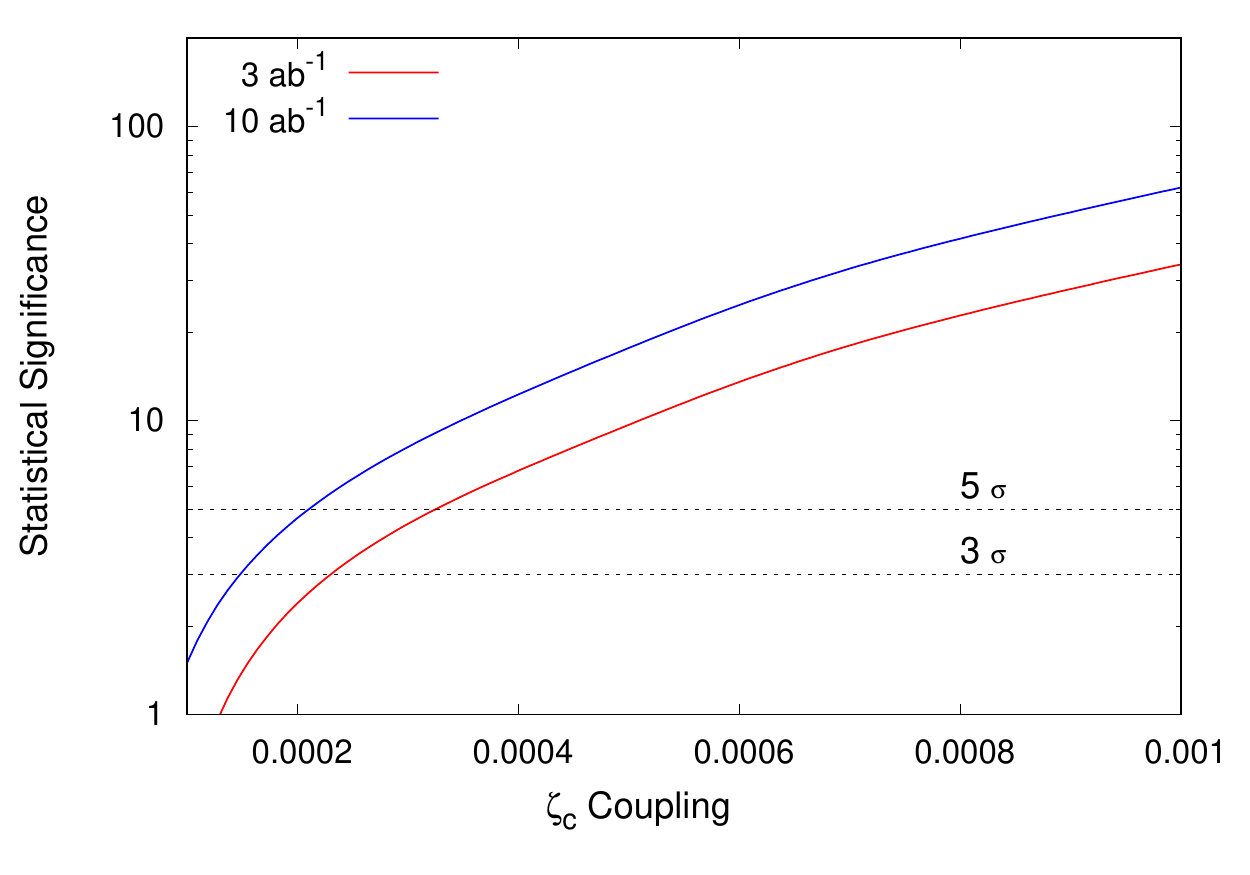}

\caption{The statistical significance (SS) as a function of the anomalous FCNC top couplings strengths after applying an optimum BDT cut value for each at integrated luminosities L$_{int}$=3 ab$^{-1}$ and  L$_{int}$=10 ab$^{-1}$ . Only one coupling ($\zeta_u$ or $\zeta_c$) at a time is varied from its SM value.  \label{SS}}
\end{figure}

The distributions of reconstructed top quark in Fig.\ref{mtop_inv} are normalized to the integrated luminosity of 100 fb$^{-1}$ and the range between 135 GeV and 195 GeV is used to calculate Statistical Significance (SS). 
Using Poisson formula for SS as
\begin{equation}
SS=\sqrt{2[(S+B_T)\ln(1+S/B_T)-S]}
\end{equation}
where $S$ and $B_T$ are the signal and total background events at a particular luminosity. The SS as a function of couplings $\zeta_u$ (on the left) and $\zeta_c$ (on the right) for $L_{int}$=3  ab$^{-1}$ and $L_{int}$=10  ab$^{-1}$ are shown in Fig.\ref{SS}. In this figure, only one coupling at a time is varied from its SM value. For integrated luminosity of 3  ab$^{-1}$, upper limit for $\zeta_u$ ($\zeta_c$) reaches to $2.309\times 10^{-4}$ ($2.313\times 10^{-4}$) at 3$\sigma$ SS value while $3.022\times 10^{-4}$($3.256\times 10^{-4}$) at 5$\sigma$. Increasing integrated luminosity to 10 ab$^{-1}$ lower the upper limit  but not drastically. 

\begin{figure}
\includegraphics[scale=1]{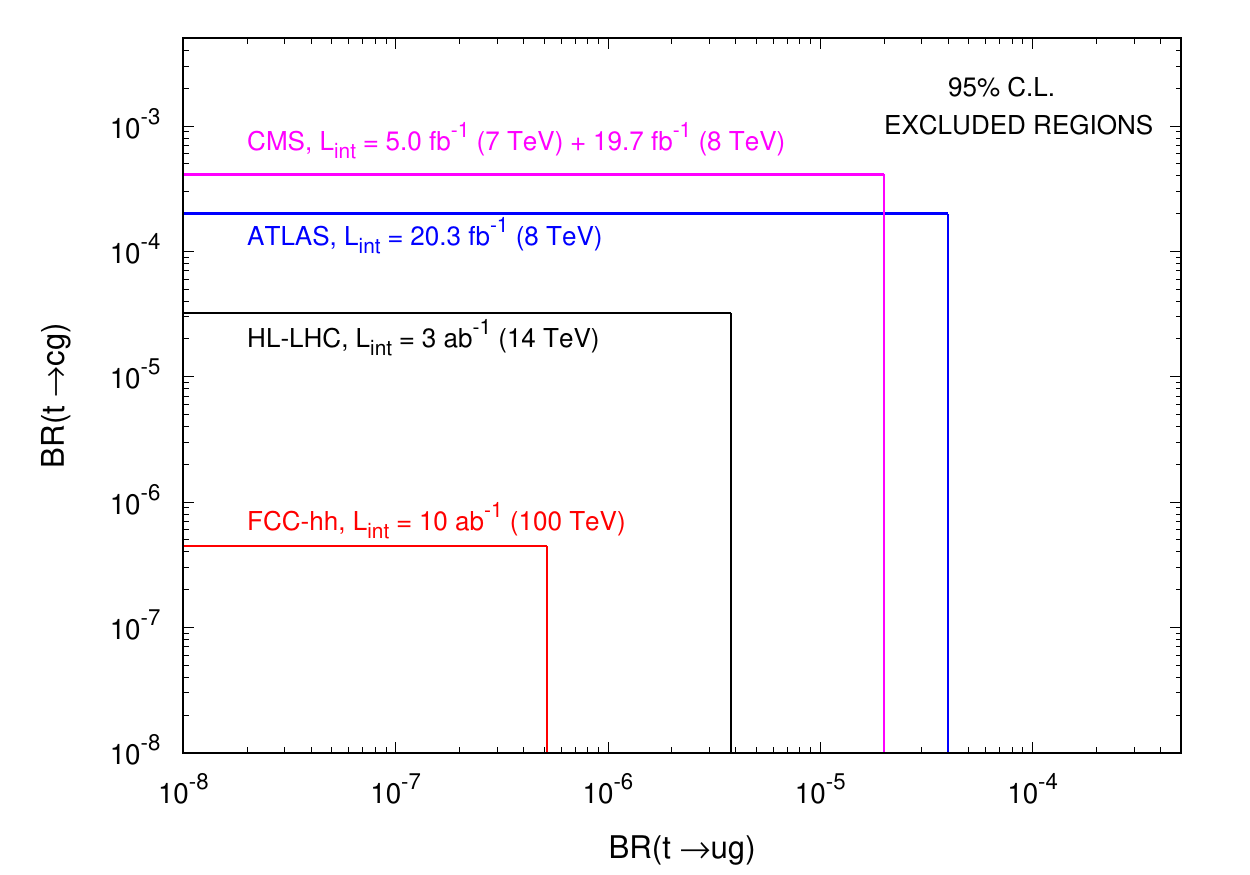}
\caption{ The current upper limits on the  $BR(t \to cg)$ versus $BR(t \to ug)$ at \% 95 CL from the recent results of ATLAS  \cite{Aad:2015gea} and CMS \cite{Khachatryan:2016sib} experiments as well as HL-LHC projection \cite{CMS:2018kwi} are compared to FCC-hh limits with $L_{int}$=10  ab$^{-1}$ at 100 TeV center of mass energy. \label{final}}
\end{figure}

The predictions for SM branching ratios of the top quark FCNC decay to gluon, photon, $Z$ or Higgs boson and up-type quarks are at the order of 10$^{-14}$. Thus, the precise measurements of these branchings can have an important role to test new physics in the top quark sector. One can express results in terms of branching ratios which can be comparable with the results of other studies as in \cite{Oyulmaz:2018irs}. In Fig.\ref{final}, the current observed \cite{Aad:2015gea, Khachatryan:2016sib} and projected upper limits \cite{CMS:2018kwi} on the branching ratios of $t\to u g$  and $t\to c g$ at 95 \% C.L. are presented and compared with the sensitivity of FCC-hh at $L_{int}$=10  ab$^{-1}$. Extracting the potential of the FCC, the sensitivity to FCNC couplings are significantly better than even the projected limits at HL-LHC \cite{CMS:2018kwi} with $L_{int}$=3  ab$^{-1}$. We obtained limits on branchings  $BR (t \to ug) =5.18\times 10^{-7}$ and $BR (t \to c g)=4.45\times 10^{-7}$ which are one order of the magnitude better than the limits for HL-LHC.
%\newpage
\section{Conclusions}
We have investigated the potential of FCC-hh collider for $qg \to l\nu b$ process at a center of mass energy of 100 TeV to set an upper limits on the anomalous top FCNC $tqg$ couplings including realistic detector effects.  The selection criteria based on Boosted Decision Trees (BDT) is used with a set of useful kinematic variables to separate signal from relevant background processes. We have shown the distributions of kinematic variables of the final state particles, transverse mass and invariant mass distribution of the top quark to define cuts for BDT training. We find the upper limits on $\zeta_u$  and $\zeta_c$ couplings at 95 \% C.L. for integrated luminosity of 3 and 10 ab$^{-1}$. With an integrated luminosity of 10 ab$^{-1}$ at the center of mass energy of 100 TeV, upper limits on  $\zeta_u$ and $\zeta_c$ are converted to branching ratios $BR (t \to ug)$ and $BR (t \to c g)$ at FCC-hh. It is found that a sensitivity of the order of 10$^{-7}$ for branching ratios at high integrated luminosity would be achievable. 
\begin{acknowledgments}
This work was supported by Turkish Atomic Energy Authority (TAEK) under the Grant No. 2018TAEK(CERN)A5.H6.F2-20. 
 \end{acknowledgments}
%\newpage

\newpage

%\newpage

\end{document}